\documentclass[10pt, conference]{IEEEtran-v17}

\pdfoutput=1
\usepackage[english]{babel}
\usepackage{setspace}

\usepackage{xspace}
\usepackage[nosort]{cite}        
\usepackage{url} 
\usepackage[cmex10]{amsmath}
\usepackage{bbm}
\usepackage{graphicx}
\usepackage{paralist}
\usepackage{fancyref}
\usepackage{pdfsync}
\usepackage[margin=2cm,top=3cm,bottom=3cm]{geometry}

\makeatletter
\def\@IEEEinterspaceratioM{0.265}
\def\@IEEEinterspaceMINratioM{0.1651}
\def\@IEEEinterspaceMAXratioM{0.38}

\def\@IEEEinterspaceratioB{0.31}
\def\@IEEEinterspaceMINratioB{0.19}
\def\@IEEEinterspaceMAXratioB{0.38}
\@IEEEtunefonts
\makeatother

\usepackage{amssymb}
\usepackage{amsfonts}
\usepackage{mathrsfs}
\usepackage{xspace}
\usepackage{bm}
\usepackage{upgreek}

\newcommand{\safemath}[2]{\newcommand{#1}{\ensuremath{#2}\xspace}}

\safemath{\bma}{\mathbf{a}}
\safemath{\bmb}{\mathbf{b}}
\safemath{\bmc}{\mathbf{c}}
\safemath{\bmd}{\mathbf{d}}
\safemath{\bme}{\mathbf{e}}
\safemath{\bmf}{\mathbf{f}}
\safemath{\bmg}{\mathbf{g}}
\safemath{\bmh}{\mathbf{h}}
\safemath{\bmi}{\mathbf{i}}
\safemath{\bmj}{\mathbf{j}}
\safemath{\bmk}{\mathbf{k}}
\safemath{\bml}{\mathbf{l}}
\safemath{\bmm}{\mathbf{m}}
\safemath{\bmn}{\mathbf{n}}
\safemath{\bmo}{\mathbf{o}}
\safemath{\bmp}{\mathbf{p}}
\safemath{\bmq}{\mathbf{q}}
\safemath{\bmr}{\mathbf{r}}
\safemath{\bms}{\mathbf{s}}
\safemath{\bmt}{\mathbf{t}}
\safemath{\bmu}{\mathbf{u}}
\safemath{\bmv}{\mathbf{v}}
\safemath{\bmw}{\mathbf{w}}
\safemath{\bmx}{\mathbf{x}}
\safemath{\bmy}{\mathbf{y}}
\safemath{\bmz}{\mathbf{z}}
\safemath{\bmzero}{\mathbf{0}}
\safemath{\bmone}{\mathbf{1}}

\bmdefine{\biad}{a}
\bmdefine{\bibd}{b}
\bmdefine{\bicd}{c}
\bmdefine{\bidd}{d}
\bmdefine{\bied}{e}
\bmdefine{\bifd}{f}
\bmdefine{\bigd}{g}
\bmdefine{\bihd}{h}
\bmdefine{\biid}{i}
\bmdefine{\bijd}{j}
\bmdefine{\bikd}{k}
\bmdefine{\bild}{l}
\bmdefine{\bimd}{m}
\bmdefine{\bind}{n}
\bmdefine{\biod}{o}
\bmdefine{\bipd}{p}
\bmdefine{\biqd}{q}
\bmdefine{\bird}{r}
\bmdefine{\bisd}{s}
\bmdefine{\bitd}{t}
\bmdefine{\biud}{u}
\bmdefine{\bivd}{v}
\bmdefine{\biwd}{w}
\bmdefine{\bixd}{x}
\bmdefine{\biyd}{y}
\bmdefine{\bizd}{z}

\bmdefine{\bixid}{\xi}
\bmdefine{\bilambdad}{\lambda}
\bmdefine{\bimud}{\mu}
\bmdefine{\bithetad}{\theta}
\bmdefine{\biphid}{\phi}

\safemath{\bmia}{\biad}
\safemath{\bmib}{\bibd}
\safemath{\bmic}{\bicd}
\safemath{\bmid}{\bidd}
\safemath{\bmie}{\bied}
\safemath{\bmif}{\bifd}
\safemath{\bmig}{\bigd}
\safemath{\bmih}{\bihd}
\safemath{\bmii}{\biid}
\safemath{\bmij}{\bijd}
\safemath{\bmik}{\bikd}
\safemath{\bmil}{\bild}
\safemath{\bmim}{\bimd}
\safemath{\bmin}{\bind}
\safemath{\bmio}{\biod}
\safemath{\bmip}{\bipd}
\safemath{\bmiq}{\biqd}
\safemath{\bmir}{\bird}
\safemath{\bmis}{\bisd}
\safemath{\bmit}{\bitd}
\safemath{\bmiu}{\biud}
\safemath{\bmiv}{\bivd}
\safemath{\bmiw}{\biwd}
\safemath{\bmix}{\bixd}
\safemath{\bmiy}{\biyd}
\safemath{\bmiz}{\bizd}

\safemath{\bmxi}{\bixid}
\safemath{\bmlambda}{\bilambdad}
\safemath{\bmmu}{\bimud}
\safemath{\bmtheta}{\bithetad}
\safemath{\bmphi}{\biphid}

\safemath{\bA}{\mathbf{A}}
\safemath{\bB}{\mathbf{B}}
\safemath{\bC}{\mathbf{C}}
\safemath{\bD}{\mathbf{D}}
\safemath{\bE}{\mathbf{E}}
\safemath{\bF}{\mathbf{F}}
\safemath{\bG}{\mathbf{G}}
\safemath{\bH}{\mathbf{H}}
\safemath{\bI}{\mathbf{I}}
\safemath{\bJ}{\mathbf{J}}
\safemath{\bK}{\mathbf{K}}
\safemath{\bL}{\mathbf{L}}
\safemath{\bM}{\mathbf{M}}
\safemath{\bN}{\mathbf{N}}
\safemath{\bO}{\mathbf{O}}
\safemath{\bP}{\mathbf{P}}
\safemath{\bQ}{\mathbf{Q}}
\safemath{\bR}{\mathbf{R}}
\safemath{\bS}{\mathbf{S}}
\safemath{\bT}{\mathbf{T}}
\safemath{\bU}{\mathbf{U}}
\safemath{\bV}{\mathbf{V}}
\safemath{\bW}{\mathbf{W}}
\safemath{\bX}{\mathbf{X}}
\safemath{\bY}{\mathbf{Y}}
\safemath{\bZ}{\mathbf{Z}}

\safemath{\bZero}{\mathbf{0}}
\safemath{\bOne}{\mathbf{1}}
\safemath{\bDelta}{\mathbf{\Delta}}
\safemath{\bLambda}{\mathbf{\UpLambda}}
\safemath{\bPhi}{\mathbf{\Upphi}}
\safemath{\bSigma}{\mathbf{\Upsigma}}
\safemath{\bOmega}{\mathbf{\Upomega}}
\safemath{\bTheta}{\mathbf{\Uptheta}}

\bmdefine{\biAd}{A}
\bmdefine{\biBd}{B}
\bmdefine{\biCd}{C}
\bmdefine{\biDd}{D}
\bmdefine{\biEd}{E}
\bmdefine{\biFd}{F}
\bmdefine{\biGd}{G}
\bmdefine{\biHd}{H}
\bmdefine{\biId}{I}
\bmdefine{\biJd}{J}
\bmdefine{\biKd}{K}
\bmdefine{\biLd}{L}
\bmdefine{\biMd}{M}
\bmdefine{\biOd}{N}
\bmdefine{\biPd}{O}
\bmdefine{\biQd}{P}
\bmdefine{\biRd}{R}
\bmdefine{\biSd}{S}
\bmdefine{\biTd}{T}
\bmdefine{\biUd}{U}
\bmdefine{\biVd}{V}
\bmdefine{\biWd}{W}
\bmdefine{\biXd}{X}
\bmdefine{\biYd}{Y}
\bmdefine{\biZd}{Z}

\bmdefine{\biDelta}{\Delta}
\bmdefine{\biLambda}{\Lambda}
\bmdefine{\biPhi}{\Phi}
\bmdefine{\biSigma}{\Sigma}
\bmdefine{\biOmega}{\Omega}
\bmdefine{\biTheta}{\Theta}

\safemath{\bimA}{\biAd}
\safemath{\bimB}{\biBd}
\safemath{\bimC}{\biCd}
\safemath{\bimD}{\biDd}
\safemath{\bimE}{\biEd}
\safemath{\bimF}{\biFd}
\safemath{\bimG}{\biGd}
\safemath{\bimH}{\biHd}
\safemath{\bimI}{\biId}
\safemath{\bimJ}{\biJd}
\safemath{\bimK}{\biKd}
\safemath{\bimL}{\biLd}
\safemath{\bimM}{\biMd}
\safemath{\bimN}{\biNd}
\safemath{\bimO}{\biOd}
\safemath{\bimP}{\biPd}
\safemath{\bimQ}{\biQd}
\safemath{\bimR}{\biRd}
\safemath{\bimS}{\biSd}
\safemath{\bimT}{\biTd}
\safemath{\bimU}{\biUd}
\safemath{\bimV}{\biVd}
\safemath{\bimW}{\biWd}
\safemath{\bimX}{\biXd}
\safemath{\bimY}{\biYd}
\safemath{\bimZ}{\biZd}

\safemath{\bimDelta}{\biDelta}
\safemath{\bimLambda}{\biLambda}
\safemath{\bimPhi}{\biPhi}
\safemath{\bimSigma}{\biSigma}
\safemath{\bimOmega}{\biOmega}
\safemath{\bimTheta}{\biTheta}

\safemath{\setA}{\mathcal{A}}
\safemath{\setB}{\mathcal{B}}
\safemath{\setC}{\mathcal{C}}
\safemath{\setD}{\mathcal{D}}
\safemath{\setE}{\mathcal{E}}
\safemath{\setF}{\mathcal{F}}
\safemath{\setG}{\mathcal{G}}
\safemath{\setH}{\mathcal{H}}
\safemath{\setI}{\mathcal{I}}
\safemath{\setJ}{\mathcal{J}}
\safemath{\setK}{\mathcal{K}}
\safemath{\setL}{\mathcal{L}}
\safemath{\setM}{\mathcal{M}}
\safemath{\setN}{\mathcal{N}}
\safemath{\setO}{\mathcal{O}}
\safemath{\setP}{\mathcal{P}}
\safemath{\setQ}{\mathcal{Q}}
\safemath{\setR}{\mathcal{R}}
\safemath{\setS}{\mathcal{S}}
\safemath{\setT}{\mathcal{T}}
\safemath{\setU}{\mathcal{U}}
\safemath{\setV}{\mathcal{V}}
\safemath{\setW}{\mathcal{W}}
\safemath{\setX}{\mathcal{X}}
\safemath{\setY}{\mathcal{Y}}
\safemath{\setZ}{\mathcal{Z}}
\safemath{\emptySet}{\varnothing}

\safemath{\colA}{\mathscr{A}}
\safemath{\colB}{\mathscr{B}}
\safemath{\colC}{\mathscr{C}}
\safemath{\colD}{\mathscr{D}}
\safemath{\colE}{\mathscr{E}}
\safemath{\colF}{\mathscr{F}}
\safemath{\colG}{\mathscr{G}}
\safemath{\colH}{\mathscr{H}}
\safemath{\colI}{\mathscr{I}}
\safemath{\colJ}{\mathscr{J}}
\safemath{\colK}{\mathscr{K}}
\safemath{\colL}{\mathscr{L}}
\safemath{\colM}{\mathscr{M}}
\safemath{\colN}{\mathscr{N}}
\safemath{\colO}{\mathscr{O}}
\safemath{\colP}{\mathscr{P}}
\safemath{\colQ}{\mathscr{Q}}
\safemath{\colR}{\mathscr{R}}
\safemath{\colS}{\mathscr{S}}
\safemath{\colT}{\mathscr{T}}
\safemath{\colU}{\mathscr{U}}
\safemath{\colV}{\mathscr{V}}
\safemath{\colW}{\mathscr{W}}
\safemath{\colX}{\mathscr{X}}
\safemath{\colY}{\mathscr{Y}}
\safemath{\colZ}{\mathscr{Z}}

\safemath{\opA}{\mathbb{A}}
\safemath{\opB}{\mathbb{B}}
\safemath{\opC}{\mathbb{C}}
\safemath{\opD}{\mathbb{D}}
\safemath{\opE}{\mathbb{E}}
\safemath{\opF}{\mathbb{F}}
\safemath{\opG}{\mathbb{G}}
\safemath{\opH}{\mathbb{H}}
\safemath{\opI}{\mathbb{I}}
\safemath{\opJ}{\mathbb{J}}
\safemath{\opK}{\mathbb{K}}
\safemath{\opL}{\mathbb{L}}
\safemath{\opM}{\mathbb{M}}
\safemath{\opN}{\mathbb{N}}
\safemath{\opO}{\mathbb{O}}
\safemath{\opP}{\mathbb{P}}
\safemath{\opQ}{\mathbb{Q}}
\safemath{\opR}{\mathbb{R}}
\safemath{\opS}{\mathbb{S}}
\safemath{\opT}{\mathbb{T}}
\safemath{\opU}{\mathbb{U}}
\safemath{\opV}{\mathbb{V}}
\safemath{\opW}{\mathbb{W}}
\safemath{\opX}{\mathbb{X}}
\safemath{\opY}{\mathbb{Y}}
\safemath{\opZ}{\mathbb{Z}}
\safemath{\opZero}{\mathbb{O}}
\safemath{\identityop}{\opI}

\safemath{\veca}{\bma}
\safemath{\vecb}{\bmb}
\safemath{\vecc}{\bmc}
\safemath{\vecd}{\bmd}
\safemath{\vece}{\bme}
\safemath{\vecf}{\bmf}
\safemath{\vecg}{\bmg}
\safemath{\vech}{\bmh}
\safemath{\veci}{\bmi}
\safemath{\vecj}{\bmj}
\safemath{\veck}{\bmk}
\safemath{\vecl}{\bml}
\safemath{\vecm}{\bmm}
\safemath{\vecn}{\bmn}
\safemath{\veco}{\bmo}
\safemath{\vecp}{\bmp}
\safemath{\vecq}{\bmq}
\safemath{\vecr}{\bmr}
\safemath{\vecs}{\bms}
\safemath{\vect}{\bmt}
\safemath{\vecu}{\bmu}
\safemath{\vecv}{\bmv}
\safemath{\vecw}{\bmw}
\safemath{\vecx}{\bmx}
\safemath{\vecy}{\bmy}
\safemath{\vecz}{\bmz}

\safemath{\veczero}{\bmzero}
\safemath{\vecone}{\bmone}
\safemath{\vecxi}{\bmxi}
\safemath{\veclambda}{\bmlambda}
\safemath{\vecmu}{\bmmu}
\safemath{\vectheta}{\bmtheta}
\safemath{\vecphi}{\bmphi}

\safemath{\matA}{\bA}
\safemath{\matB}{\bB}
\safemath{\matC}{\bC}
\safemath{\matD}{\bD}
\safemath{\matE}{\bE}
\safemath{\matF}{\bF}
\safemath{\matG}{\bG}
\safemath{\matH}{\bH}
\safemath{\matI}{\bI}
\safemath{\matJ}{\bJ}
\safemath{\matK}{\bK}
\safemath{\matL}{\bL}
\safemath{\matM}{\bM}
\safemath{\matN}{\bN}
\safemath{\matO}{\bO}
\safemath{\matP}{\bP}
\safemath{\matQ}{\bQ}
\safemath{\matR}{\bR}
\safemath{\matS}{\bS}
\safemath{\matT}{\bT}
\safemath{\matU}{\bU}
\safemath{\matV}{\bV}
\safemath{\matW}{\bW}
\safemath{\matX}{\bX}
\safemath{\matY}{\bY}
\safemath{\matZ}{\bZ}
\safemath{\matzero}{\bmzero}

\safemath{\matDelta}{\bDelta}
\safemath{\matLambda}{\bLambda}
\safemath{\matPhi}{\bPhi}
\safemath{\matSigma}{\bSigma}
\safemath{\matOmega}{\bOmega}
\safemath{\matTheta}{\bTheta}

\safemath{\matidentity}{\matI}
\safemath{\matone}{\matO}

\safemath{\rnda}{A}
\safemath{\rndb}{B}
\safemath{\rndc}{C}
\safemath{\rndd}{D}
\safemath{\rnde}{E}
\safemath{\rndf}{F}
\safemath{\rndg}{G}
\safemath{\rndh}{H}
\safemath{\rndi}{I}
\safemath{\rndj}{J}
\safemath{\rndk}{K}
\safemath{\rndl}{L}
\safemath{\rndm}{M}
\safemath{\rndn}{N}
\safemath{\rndo}{O}
\safemath{\rndp}{P}
\safemath{\rndq}{Q}
\safemath{\rndr}{R}
\safemath{\rnds}{S}
\safemath{\rndt}{T}
\safemath{\rndu}{U}
\safemath{\rndv}{V}
\safemath{\rndw}{W}
\safemath{\rndx}{X}
\safemath{\rndy}{Y}
\safemath{\rndz}{Z}

\safemath{\rveca}{\bimA}
\safemath{\rvecb}{\bimB}
\safemath{\rvecc}{\bimC}
\safemath{\rvecd}{\bimD}
\safemath{\rvece}{\bimE}
\safemath{\rvecf}{\bimF}
\safemath{\rvecg}{\bimG}
\safemath{\rvech}{\bimH}
\safemath{\rveci}{\bimI}
\safemath{\rvecj}{\bimJ}
\safemath{\rveck}{\bimK}
\safemath{\rvecl}{\bimL}
\safemath{\rvecm}{\bimM}
\safemath{\rvecn}{\bimN}
\safemath{\rveco}{\bomO}
\safemath{\rvecp}{\bimP}
\safemath{\rvecq}{\bimQ}
\safemath{\rvecr}{\bimR}
\safemath{\rvecs}{\bimS}
\safemath{\rvect}{\bimT}
\safemath{\rvecu}{\bimU}
\safemath{\rvecv}{\bimV}
\safemath{\rvecw}{\bimW}
\safemath{\rvecx}{\bimX}
\safemath{\rvecy}{\bimY}
\safemath{\rvecz}{\bimZ}

\safemath{\rvecxi}{\bmxi}
\safemath{\rveclambda}{\bmlambda}
\safemath{\rvecmu}{\bmmu}
\safemath{\rvectheta}{\bmtheta}
\safemath{\rvecphi}{\bmphi}

\safemath{\rmatA}{\bimA}
\safemath{\rmatB}{\bimB}
\safemath{\rmatC}{\bimC}
\safemath{\rmatD}{\bimD}
\safemath{\rmatE}{\bimE}
\safemath{\rmatF}{\bimF}
\safemath{\rmatG}{\bimG}
\safemath{\rmatH}{\bimH}
\safemath{\rmatI}{\bimI}
\safemath{\rmatJ}{\bimJ}
\safemath{\rmatK}{\bimK}
\safemath{\rmatL}{\bimL}
\safemath{\rmatM}{\bimM}
\safemath{\rmatN}{\bimN}
\safemath{\rmatO}{\bimO}
\safemath{\rmatP}{\bimP}
\safemath{\rmatQ}{\bimQ}
\safemath{\rmatR}{\bimR}
\safemath{\rmatS}{\bimS}
\safemath{\rmatT}{\bimT}
\safemath{\rmatU}{\bimU}
\safemath{\rmatV}{\bimV}
\safemath{\rmatW}{\bimW}
\safemath{\rmatX}{\bimX}
\safemath{\rmatY}{\bimY}
\safemath{\rmatZ}{\bimZ}

\safemath{\rmatDelta}{\bimDelta}
\safemath{\rmatLambda}{\bimLambda}
\safemath{\rmatPhi}{\bimPhi}
\safemath{\rmatSigma}{\bimSigma}
\safemath{\rmatOmega}{\bimOmega}
\safemath{\rmatTheta}{\bimTheta}

\usepackage{amssymb}
\usepackage{amsfonts}
\usepackage{mathrsfs}
\usepackage{xspace}
\usepackage{bm}
\usepackage{fancyref}
\usepackage{textcomp}

\newenvironment{textbmatrix}{	\setlength{\arraycolsep}{2.5pt}%
								\big[\begin{matrix}}{\end{matrix}\big]%
								\raisebox{0.08ex}{\vphantom{M}}}

\def\be{\begin{equation}}
\def\ee{\end{equation}}
\def\een{\nonumber \end{equation}}
\def\mat{\begin{bmatrix}}
\def\emat{\end{bmatrix}}
\def\btm{\begin{textbmatrix}}
\def\etm{\end{textbmatrix}}

\def\ba#1\ea{\begin{align}#1\end{align}}
\def\bas#1\eas{\begin{align*}#1\end{align*}}
\def\bs#1\es{\begin{split}#1\end{split}} 
\def\bg#1\eg{\begin{gather}#1\end{gather}}
\def\bml#1\eml{\begin{multline}#1\end{multline}}
\def\bi#1\ei{\begin{itemize}#1\end{itemize}}

\newcommand{\lefto}{\mathopen{}\left}

\DeclareMathOperator{\rank}{rank}			
\DeclareMathOperator{\Prob}{\opP}			
\DeclareMathOperator{\Exop}{\opE}			

\newcommand{\abs}[1]{\lefto\lvert#1\right\rvert}		

\newcommand{\vecnorm}[1]{\lefto\lVert#1\right\rVert}		
\newcommand{\herm}[1]{\ensuremath{#1^{H}}} 	

\safemath{\dirac}{\delta}					
\safemath{\krond}{\dirac}					

\safemath{\upto}{\uparrow}
\safemath{\downto}{\downarrow}
\safemath{\iu}{j}							
\safemath{\ev}{\lambda}						
\safemath{\hilseqspace}{l^{2}}				
\newcommand{\banachfunspace}[1]{\setL^{#1}}	
\safemath{\hilfunspace}{\banachfunspace{2}}	

\safemath{\SNR}{\text{\sc snr}} 				
\safemath{\No}{N_0}							
\safemath{\Es}{E_s}							
\safemath{\Eb}{E_b}							
\safemath{\EbNo}{\frac{\Eb}{\No}}
\safemath{\EsNo}{\frac{\Es}{\No}}

\DeclareMathOperator{\CHop}{\ensuremath{\opH}} 
\safemath{\tvir}{\rndh_{\CHop}}				
\safemath{\tvtf}{\rndl_{\CHop}}				
\safemath{\spf}{\rnds_{\CHop}}				
\safemath{\bff}{H_{\CHop}}					

\safemath{\ircf}{r_{h}}						
\safemath{\tftvcf}{r_{s}}					
\safemath{\tfcf}{r_{l}}						
\safemath{\bfcf}{r_{H}}						

\safemath{\tcorr}{c_h}						
\safemath{\scf}{c_{s}}						
\safemath{\tfcorr}{c_{l}}					
\safemath{\fcorr}{c_{H}}						

\safemath{\mi}{I}							
\safemath{\capacity}{C}						

\safemath{\normal}{\mathcal{N}}			
\safemath{\jpg}{\mathcal{CN}}			
\safemath{\mchain}{\leftrightarrow}		

\safemath{\dB}{\,\mathrm{dB}}
\safemath{\dBm}{\,\mathrm{dBm}}
\safemath{\Hz}{\,\mathrm{Hz}}
\safemath{\kHz}{\,\mathrm{kHz}}
\safemath{\MHz}{\,\mathrm{MHz}}
\safemath{\GHz}{\,\mathrm{GHz}}
\safemath{\s}{\,\mathrm{s}}
\safemath{\ms}{\,\mathrm{ms}}
\safemath{\mus}{\,\mathrm{\text{\textmu}s}}
\safemath{\ns}{\,\mathrm{ns}}
\safemath{\ps}{\,\mathrm{ps}}
\safemath{\meter}{\,\mathrm{m}}
\safemath{\mm}{\,\mathrm{mm}}
\safemath{\cm}{\,\mathrm{cm}}
\safemath{\m}{\,\mathrm{m}}
\safemath{\W}{\,\mathrm{W}}
\safemath{\mW}{\, \mathrm{mW}}
\safemath{\J}{\,\mathrm{J}}
\safemath{\K}{\,\mathrm{K}}
\safemath{\bit}{\,\mathrm{bit}}
\safemath{\nat}{\,\mathrm{nat}}

\safemath{\define}{=}			

\safemath{\equivalent}{\sim}
\safemath{\distas}{\sim}					
\safemath{\sdiff}{\Delta}				

\safemath{\reals}{\mathbb{R}}
\safemath{\positivereals}{\reals_{+}}
\safemath{\integers}{\mathbb{Z}}
\safemath{\posint}{\integers_{+}}
\safemath{\naturals}{\mathbb{N}}
\safemath{\posnaturals}{\naturals_{+}}
\safemath{\complexset}{\mathbb{C}}
\safemath{\rationals}{\mathbb{Q}}

\newcommand*{\fancyrefapplabelprefix}{app}		
\newcommand*{\fancyrefthmlabelprefix}{thm}		
\newcommand*{\fancyreflemlabelprefix}{lem}		
\newcommand*{\fancyrefcorlabelprefix}{cor}		
\newcommand*{\fancyrefdeflabelprefix}{def}		
\newcommand*{\fancyrefproplabelprefix}{prop}		
\frefformat{vario}{\fancyrefseclabelprefix}{Section~#1}
\frefformat{vario}{\fancyrefthmlabelprefix}{Theorem~#1}
\frefformat{vario}{\fancyreflemlabelprefix}{Lemma~#1}
\frefformat{vario}{\fancyrefcorlabelprefix}{Corollary~#1}
\frefformat{vario}{\fancyrefdeflabelprefix}{Definition~#1}
\frefformat{vario}{\fancyreffiglabelprefix}{Fig.~#1}
\frefformat{vario}{\fancyrefapplabelprefix}{Appendix~#1}
\frefformat{vario}{\fancyrefeqlabelprefix}{(#1)}
\frefformat{vario}{\fancyrefproplabelprefix}{Property~#1}

 \newcommand{\trm}{\textrm}
 \newcommand{\dummyrel}[1]{\mathrel{\hphantom{#1}}\strut\mskip-\medmuskip}
 \newtheorem{thm}{Theorem}

 \newtheorem{lem}{Lemma}

\safemath{\dict}{\matD}
\safemath{\inputdim}{N}		
\safemath{\outputdim}{M}		
\safemath{\sparsity}{S}	
\safemath{\inputdimA}{{N_a}}	
\safemath{\inputdimB}{{N_b}}	
\safemath{\elemA}{{n_a}}	
\safemath{\elemB}{{n_b}}	
\safemath{\resA}{\matR_a}	
\safemath{\resB}{\matR_b}	
\safemath{\subD}{\matS} 
\safemath{\subA}{\matS_a} 
\safemath{\subB}{\matS_b} 
\safemath{\dicta}{\matA} 	
\safemath{\dictb}{\matB} 	
\safemath{\hollowS}{H}
\safemath{\hollowA}{H_a}
\safemath{\hollowB}{H_b}
\safemath{\cross}{Z}
\safemath{\coh}{d}			
\safemath{\coha}{a}			
\safemath{\cohb}{b}			
\safemath{\dictset}{\setD}	
\safemath{\dictsetp}{\dictset(\coh,\coha,\cohb)}	
\safemath{\dictsetgen}{\dictset_\text{gen}}
\safemath{\dictsetgenp}{\dictsetgen(\coh)}
\safemath{\dictsetonb}{\dictset_\text{onb}}
\safemath{\dictsetonbp}{\dictsetonb(\coh)}

\safemath{\na}{n_a}			
\safemath{\nb}{n_b}			
\safemath{\coeffa}{p_i}	
\safemath{\coeffb}{q_j}	
\safemath{\seta}{\setP}		
\safemath{\setb}{\setQ}     
\safemath{\setw}{\setW}	
\safemath{\setz}{\setZ}	
\safemath{\cola}{\veca}		
\safemath{\colb}{\vecb}		
\safemath{\cold}{\vecd}		
\safemath{\inputvec}{\vecx} 	
\safemath{\inputvecel}{x}
\safemath{\inputveca}{\vecx_a}
\safemath{\inputvecb}{\vecx_b}
\safemath{\outputvec}{\vecy}	
\safemath{\lambdamin}{\lambda_{\mathrm{min}}}

\newcommand{\normtwo}[1]{\vecnorm{#1}_2}
\newcommand{\normone}[1]{\vecnorm{#1}_1}
\newcommand{\normzero}[1]{\vecnorm{#1}_0}

\newcommand{\spectralnorm}[1]{\vecnorm{#1}}
\safemath{\elltwo}{\ell_2}
\safemath{\ellone}{\ell_1}
\safemath{\ellzero}{\ell_0}
\safemath{\ellinf}{\ell_\infty}
\safemath{\licard}{Z(\dict)}
\safemath{\xsol}{\hat{x}}
\safemath{\xbord}{x_b}		
\safemath{\xstat}{x_s}		
\safemath{\xstatLone}{\tilde{x}_s}
\safemath{\order}{\mathcal{O}} 
\safemath{\scales}{\Theta} 
\safemath{\thlone}{\kappa(\coh,\cohb)} 
\safemath{\constoneA}{\delta} 
\safemath{\constoneB}{\epsilon} 
\safemath{\nlarge}{L}				   
\safemath{\sumlarge}{S_\nlarge}
\safemath{\maxlarger}{P_\nlarge}	   
\safemath{\Pzero}{\textrm{P0}}	
\safemath{\Pone}{\textrm{P1}}
\safemath{\vecfir}{\vecw}			 
\safemath{\vecsec}{\vecz}
\safemath{\elvecfir}{w}              
\safemath{\elvecsec}{z}				 
\safemath{\nlargefir}{n}
\safemath{\normout}{\gamma}
\safemath{\auxfun}{h}
\safemath{\supp}{\textrm{supp}}

\begin{document}
\renewcommand{\baselinestretch}{0.99}\small\normalsize
	%
	%
\title{Where is Randomness Needed to Break the Square-Root Bottleneck?}

\author{
\IEEEauthorblockN{Patrick Kuppinger, Giuseppe Durisi, and Helmut B\"olcskei\\}
\IEEEauthorblockA{ETH Zurich, 8092 Zurich, Switzerland\\
E-mail: \{patricku, gdurisi, boelcskei\}@nari.ee.ethz.ch\\} 
}

\maketitle
\begin{abstract}
As shown by Tropp, 2008, for the concatenation of two orthonormal bases (ONBs), breaking the square-root bottleneck in compressed sensing does not require randomization over all the positions of the nonzero entries of the sparse coefficient vector. Rather the positions corresponding to one of the two ONBs can be chosen arbitrarily. The two-ONB structure is, however, restrictive and does not reveal the property that is responsible for allowing to break the bottleneck with reduced randomness. For general dictionaries we show that if a sub-dictionary with small enough coherence and large enough cardinality can be isolated, the bottleneck can be broken under the same probabilistic model on the sparse coefficient vector as in the two-ONB case.
%
\end{abstract}
\section{Introduction}
The central idea underlying compressed sensing (CS) is to recover a sparse signal from as few non-adaptive linear measurements as possible~\cite{candes2006c,donoho2006}. Given the measurement outcome $\outputvec\in\complexset^\outputdim$ and the measurement matrix $\dict\in\complexset^{\outputdim\times\inputdim}$ ($\outputdim\leq\inputdim$),  often referred to as dictionary,\footnote{Throughout the paper, we assume that the columns $\cold_i$ of \dict have unit \elltwo-norm, i.e., $\normtwo{\cold_i}=1$ for $i=1,\ldots,\inputdim$.} we want to find the sparsest coefficient vector $\inputvec\in\complexset^\inputdim$ that is consistent with the measurement outcome, i.e., that satisfies $\outputvec=\dict\inputvec$.  This problem can be formalized as follows:
\be
	(\Pzero)\quad \text{find\:\:} \arg\min \normzero{\inputvec}\quad\trm{subject to } \outputvec=\dict\inputvec.
\een
Here, $\normzero{\inputvec}$ denotes the number of nonzero entries of the vector \inputvec. Unfortunately, solving (\Pzero) for practically relevant problem sizes $\inputdim,\outputdim$ is infeasible as it requires a combinatorial search. Instead, the CS literature has focused on the convex relaxation of (\Pzero), i.e., on the following \ellone-minimization problem:
\be
	(\Pone)\quad \text{find\:\:}\arg\min\normone{\inputvec}\quad\trm{subject to } \outputvec=\dict\inputvec
\een
commonly referred to as \emph{basis pursuit} (BP)~\cite{chen1998,donoho2001,donoho2002,gribonval2003,elad2002,tropp2004}. Here, $\normone{\inputvec}\triangleq\sum_{i=1}^\inputdim\abs{x_i}$ denotes the \ellone-norm of \inputvec. Since (\Pone) can be cast as a linear program (in the real case) or a second-order cone program (in the complex case), it can be solved more efficiently than (\Pzero).

It is now natural to ask under which conditions the solutions of (\Pzero) and (\Pone) are unique and coincide. 
A sufficient condition for this to happen\footnote{In the remainder of the paper, whenever we speak of a vector \inputvec, we implicitly assume that this vector is consistent with the observation \outputvec, i.e., it satisfies $\outputvec=\dict\inputvec$.}~\cite{donoho2001,donoho2002,gribonval2003} is $\normzero{\inputvec}<\sparsity$, where the \emph{sparsity threshold} $\sparsity=(1+1/\coh)/2$ depends on the dictionary \emph{coherence} $\coh = \max_{i\neq j}\abs{\herm{\cold_i}\cold_j}$.
%
%
Sparsity thresholds \sparsity larger than $(1+1/\coh)/2$ can be established if more information on the dictionary is available~\cite{gribonval2003,elad2002,tropp2004,kuppinger2009b}, e.g., if the dictionary consists of the concatenation of two or more orthonormal bases (ONBs), or---more generally---if a sufficiently large sub-dictionary with coherence much smaller than \coh can be isolated~\cite{kuppinger2009b}. We emphasize that the results in~\cite{donoho2001,donoho2002,gribonval2003,elad2002,tropp2004,kuppinger2009b} apply to \emph{all} vectors \inputvec with $\normzero{\inputvec}<\sparsity$---irrespective of the positions and the values of the nonzero entries of \inputvec.

The line of work presented in~\cite{donoho2001,donoho2002,gribonval2003,elad2002,tropp2004,kuppinger2009b} leads to sparsity thresholds \sparsity that are on the order of $1/\coh$. From the Welch lower bound~\cite{welch1974lower} 
\be
	\coh\geq\sqrt{(\inputdim-\outputdim)/[\outputdim(\inputdim-1)]}
\een
we can conclude that the thresholds in~\cite{donoho2001,donoho2002,gribonval2003,elad2002,tropp2004,kuppinger2009b} are at best on the order of $\sqrt{\outputdim}$ (for $\inputdim\gg\outputdim$).
This scaling behavior is sometimes referred to as the \emph{square-root bottleneck}.  
A better scaling behavior can be obtained by asking for sparsity thresholds that hold for almost all---rather than all (as in~\cite{donoho2001,donoho2002,gribonval2003,elad2002,tropp2004,kuppinger2009b})---vectors \inputvec, or, more precisely, by asking for sparsity thresholds that hold with high probability,  given a probabilistic model on \inputvec.\footnote{An alternative approach, which we do not pursue in this paper, is to introduce a probabilistic model on the dictionary \dict~\cite{candes2006c,donoho2006}.}
Following the terminology used in~\cite{candes2006b}, we refer to sparsity thresholds that hold for almost all \sparsity-sparse vectors \inputvec as \emph{robust sparsity thresholds}. 

The improvements in the scaling behavior that result from the relaxation to robust sparsity thresholds will, of course, depend on the probabilistic model on \inputvec~\cite{candes2006b,tropp2008,calderbank2009a}. A widely used probabilistic model for $n$-sparse vectors \inputvec is to choose the positions of the $n$ nonzero entries (i.e., the \emph{sparsity pattern}) of \inputvec uniformly at random among all possible $\binom{\inputdim}{n}$ support sets of cardinality $n$. The values of these nonzero entries of \inputvec are drawn from a continuous probability distribution, with the additional constraint that their phases are i.i.d. and uniformly distributed on $[0,2\pi)$~\cite{candes2006b,tropp2008}.
For this probabilistic model it is shown in~\cite{tropp2008} that the square-root bottleneck can be broken. More specifically, the main result in~\cite{tropp2008} states that, assuming a dictionary with coherence on the order of  $1/\sqrt{\outputdim}$, a robust sparsity threshold on the order of $\outputdim/(\log\inputdim)$ can be obtained.
Put differently, this result shows that to recover almost all vectors \inputvec with  \sparsity nonzero entries, the required number of non-adaptive linear measurements $\outputdim$  is  (order-wise) $\sparsity\log\inputdim$ instead of $\sparsity^2$.

Remarkably, for dictionaries that consist of the concatenation of two ONBs, robust sparsity thresholds on the order of $\outputdim/(\log\inputdim)$ can be obtained with reduced randomness as compared to the case of general dictionaries. Specifically, it was found in~\cite{candes2006b,tropp2008} that it suffices to pick the positions of the nonzero entries of \inputvec corresponding to one of the two ONBs uniformly at random, while the positions of the remaining nonzero entries can be chosen arbitrarily. The probabilistic model on the values of the nonzero entries of \inputvec (corresponding to \emph{both} ONBs) remains the same as for the general dictionaries considered in~\cite{tropp2008}.  
%
%


\paragraph*{Contributions}
The two-ONB result in~\cite{candes2006b,tropp2008} is interesting as it shows that one need not choose the locations of all the nonzero entries of the sparse vector randomly to break the square-root bottleneck. However, the two-ONB structure is restrictive and does not reveal which property of the dictionary is responsible for allowing to break the square-root bottleneck with reduced randomness. The two ONBs are on equal footing. 

The purpose of this paper is twofold. First, we extend the two-ONB result in~\cite{candes2006b,tropp2008} to general dictionaries. Second, by virtue of this extension, we show that---for a general dictionary \dict with low coherence \coh---the fundamental property needed to break the square-root bottleneck with reduced randomness is the presence of  a  sufficiently large sub-dictionary \dicta with coherence much smaller than \coh. The positions of the nonzero entries of \inputvec corresponding to \dicta can be chosen arbitrarily, and the positions of the remaining nonzero entries must be chosen randomly. Naturally, the larger the sub-dictionary \dicta, the more significant the reduction in randomness becomes. Randomization over the remaining part of the dictionary ensures that the sparsity patterns that cannot be recovered through BP occur with small enough probability. More formally, we prove the following result. Consider a general dictionary \dict with coherence on the order of $1/\sqrt{\outputdim}$ that contains a sub-dictionary \dicta with coherence on the order of  $(\log\inputdim)/\outputdim$ and cardinality at least on the order of $\outputdim/(\log\inputdim)$. Then, a robust sparsity threshold on the order of $\outputdim/(\log\inputdim)$ can be established---and hence the square-root bottleneck is broken---under the same probabilistic model on the vector \inputvec as in the two-ONB case, whenever the spectral norms of \dicta and of the sub-dictionary containing the remaining columns of \dict  satisfy certain technical conditions. These technical conditions are trivially satisfied, e.g., for dictionaries that consist of two tight frames.
%

%
Our analysis relies heavily on the mathematical tools developed in~\cite{tropp2008} for the two-ONB setting.
\paragraph*{Notation}
Throughout the paper, we use lowercase boldface letters for column vectors, e.g., \vecx, and uppercase boldface letters for matrices, e.g., \dict. For a given matrix \dict, we denote its conjugate transpose by $\herm{\dict}$ and $\cold_i$ stands for its $i$th column. The spectral norm of a matrix \dict is $\spectralnorm{\dict}=\sqrt{\lambda}$, where $\lambda$ is the maximum eigenvalue of $\dict^H\dict$. The minimum and maximum singular value of a matrix \dict are denoted by $\sigma_\text{min}(\dict)$ and $\sigma_\text{max}(\dict)$, respectively, $\rank(\dict)$ stands for the rank of \dict, and  $\vecnorm{\dict}_{1,2}=\max_i\{\normtwo{\cold_i}\}$. We use $\bI_n$ to denote the $n\times n$ identity matrix and $\mathbf{0}$ stands for the all-zero matrix of appropriate size. The natural logarithm is denoted as $\log$.
For two functions $f(\outputdim)$ and $g(\outputdim)$, the notation $f(\outputdim)=\order(g(\outputdim))$ means that  $\lim_{\outputdim\to\infty}\abs{f(\outputdim)}/\abs{g(\outputdim)}$ is bounded above by a finite constant, and $f(\outputdim)=\scales(g(\outputdim))$ means that there exist two positive finite constants $k_1$ and $k_2$ such that $k_1\leq\lim_{\outputdim\to\infty}\abs{f(\outputdim)}/\abs{g(\outputdim)}\leq k_2$.
Whenever we say that a vector $\inputvec\in\complexset^\inputdim$ has a \emph{randomly} chosen sparsity pattern of cardinality $n$, we mean that the support set of \inputvec is chosen uniformly at random among all $\binom{\inputdim}{n}$ possible support sets of cardinality $n$. 
\section{Brief Review of Previous Relevant Results}
Robust sparsity thresholds for dictionaries consisting of two ONBs were first obtained in~\cite{candes2006b} and later improved in~\cite{tropp2008}. In~\fref{thm:Tropp_ONB} below, we restate the result in~\cite{tropp2008} in a slightly modified form, which is better suited to draw parallels to the more general case. The theorem follows by combining Theorems D, 13, and 14 in~\cite{tropp2008}.
\begin{thm}\label{thm:Tropp_ONB}
Assume that\footnote{In~\cite{tropp2008} $\outputdim\geq3$ (and hence $\inputdim\geq6$) is assumed. However, it can be shown that $\inputdim>2$ is sufficient to establish the result.} $\inputdim>2$. Let $\dict\in\complexset^{\outputdim\times\inputdim}$ be the concatenation of two ONBs \dicta and \dictb for $\complexset^\outputdim$ (i.e., $\inputdim=2\outputdim$) and denote the coherence of \dict as \coh. Fix $s\geq1$. Let the vector $\inputvec\in\complexset^\inputdim$ have an \emph{arbitrarily} chosen sparsity pattern of \elemA nonzero entries corresponding to columns of sub-dictionary \dicta and a \emph{randomly} chosen sparsity pattern of \elemB nonzero entries corresponding to columns of sub-dictionary \dictb. Suppose that 
\be\label{eq:Tropp1}
	\elemA+\elemB<\min\lefto\{c\,\coh^{-2}/(s\log\inputdim),\coh^{-2}/2\right\}
\ee
%
where $c$ is no smaller than $0.004212$. If the values of \emph{all} nonzero entries of \inputvec are drawn from a continuous probability distribution, \inputvec is the unique solution of (\Pzero) with probability exceeding $(1-\inputdim^{-s})$. Furthermore, if \elemA and \elemB, in addition to~\eqref{eq:Tropp1}, satisfy
\be\label{eq:Tropp2}
	\elemA+\elemB\leq\coh^{-2}/[8(s+1)\log\inputdim]
\ee
and the phases of \emph{all} nonzero entries of \inputvec are i.i.d. and uniformly distributed on $[0,2\pi)$, then \inputvec is the unique solution of both (\Pzero) and (\Pone) with probability exceeding $(1-3\inputdim^{-s})$.
\end{thm}
\vspace{3mm}

\paragraph*{Interpretation of~\fref{thm:Tropp_ONB}} Assume that \dict has coherence $\coh=\order(1/\sqrt{\outputdim})$. As a consequence of~\eqref{eq:Tropp1} and~\eqref{eq:Tropp2}, \fref{thm:Tropp_ONB} establishes (under certain technical conditions on the values of the nonzero entries of \inputvec) the robust sparsity threshold\footnote{Whenever for some function $g(\outputdim,\inputdim)$ we write $\scales(g(\outputdim,\inputdim))$ or $\order(g(\outputdim,\inputdim))$, we mean that the ratio $\inputdim/\outputdim$ remains fixed while $\outputdim\to\infty$.} $\sparsity>\elemA+\elemB=\scales(\outputdim/(\log\inputdim))$. 

This result is interesting as it shows that we do not need the entire sparsity pattern of \inputvec to be chosen at random but rather the positions of the non-zero entries corresponding to one of the two ONBs can be chosen arbitrarily.

In the following section, we first present (in~\fref{thm:l0_l1_unique}) an extension of the two-ONB result in~\cite{candes2006b,tropp2008} to general dictionaries. As a consequence of~\fref{thm:l0_l1_unique}, we then establish that---for a general dictionary \dict with low coherence \coh---the fundamental property that allows to break the square-root bottleneck with reduced randomness is the presence of  a sufficiently large sub-dictionary \dicta with coherence much smaller than \coh. 
%
%
\section{Main Results}\label{sec:main}
Consider a dictionary $\dict=[\dicta\,\,\dictb]$, where the sub-dictionary \dicta has \inputdimA elements (i.e., columns) and coherence \coha and the sub-dictionary \dictb has $\inputdimB=\inputdim-\inputdimA$ elements and coherence \cohb. The set of all such dictionaries is denoted as \dictsetp. Correspondingly, we view the vector \inputvec as the concatenation of the two vectors $\inputveca\in\complexset^\inputdimA$ and $\inputvecb\in\complexset^\inputdimB$ such that $\outputvec=\dict\inputvec=\dicta\inputveca+\dictb\inputvecb$. Since \dicta and \dictb are sub-dictionaries of \dict, we have $\coha,\cohb\leq\coh$.
We now state our main result.
\begin{thm}\label{thm:l0_l1_unique}
Assume that $\inputdim>2$. Let $\dict=[\dicta\,\,\dictb]$ be a dictionary in \dictsetp. Fix $s\geq1$ and $\gamma\in[0,1]$. 
Consider a random vector $\inputvec=\lefto[\inputveca^T\,\,\inputvecb^T\right]^T$ where \inputveca has an \emph{arbitrarily} chosen sparsity pattern of cardinality \elemA such that
\be\label{eq:condA}
	6\sqrt{2}\sqrt{\elemA\coh^2s\log\inputdim}+2(\elemA-1)\coha\leq(1-\gamma) e^{-1/4}
\ee
and \inputvecb has a \emph{randomly} chosen sparsity pattern of cardinality \elemB such that 
%
\be\label{eq:condB}
	24\sqrt{\elemB\cohb^2s\log\inputdim}+\frac{4\elemB}{\inputdimB}\spectralnorm{\dictb}^2+2\sqrt{\frac{\elemB}{\inputdimB}}\spectralnorm{\dicta}\!\spectralnorm{\dictb}\leq\gamma e^{-1/4}.
\ee
%
Furthermore, assume that the total number of nonzero entries of \inputvec satisfies
\be\label{eq:cond_l0}
	\elemA+\elemB<\coh^{-2}/2.
\ee
Then, if the values of \emph{all} nonzero entries of \inputvec are drawn from a continuous probability distribution, \inputvec is the unique solution of (\Pzero) with probability exceeding $(1-\inputdim^{-s})$. Furthermore, if \elemA and \elemB, in addition to~\eqref{eq:condA}--\eqref{eq:cond_l0}, satisfy
\be\label{eq:cond_l1}
		\elemA+\elemB\leq\coh^{-2}/[8(s+1)\log\inputdim]
\ee
and the phases of \emph{all} nonzero entries of \inputvec are i.i.d. and uniformly distributed on $[0,2\pi)$, then, \inputvec is the unique solution of both (\Pzero) and (\Pone) with probability exceeding $(1-3\inputdim^{-s})$.
\end{thm}
\vspace{3mm}
\begin{IEEEproof}
The proof is based on the following lemma, which is the main technical result of this paper and whose proof can be found in \fref{app:proofs}.
\begin{lem}\label{lem:subdictionary}
Fix $s\geq1$ and $\gamma\in[0,1]$. Let \subD be a sub-dictionary of $\dict=[\dicta\,\,\dictb]\in\dictsetp$ that contains \elemA  \emph{arbitrarily} chosen columns of \dicta and \elemB  columns of \dictb chosen uniformly at \emph{random}. If \elemA and \elemB satisfy conditions~\eqref{eq:condA} and~\eqref{eq:condB}, then, the minimum singular value $\sigma_\text{min}(\subD)$ of the sub-dictionary \subD obeys
\be
	\Prob\lefto\{\sigma_\text{min}(\subD)\leq 1/\sqrt{2}\right\}\leq \inputdim^{-s}.
\een
\end{lem}
\vspace{3mm}

The proof of~\fref{thm:l0_l1_unique} is then obtained from~\fref{lem:subdictionary} and the results in~\cite{tropp2008} as follows.
The sparsity pattern of \inputvec assumed in the statement of~\fref{thm:l0_l1_unique} induces a sub-dictionary \subD of \dict containing \elemA arbitrarily chosen columns of \dicta and \elemB randomly chosen columns of \dictb. As a consequence of~\fref{lem:subdictionary}, the smallest singular value of \subD exceeds $1/\sqrt{2}$ with probability at least $(1-\inputdim^{-s})$. 
This property of the sub-dictionary \subD, together with condition~\eqref{eq:cond_l0} and the requirement that the values of \emph{all} nonzero entries of \inputvec are drawn from a  continuous probability distribution,  implies, as a consequence of~\cite[Thm. 13]{tropp2008}, that \inputvec is the unique solution of (\Pzero) with probability at least $(1-\inputdim^{-s})$.
If, in addition, condition~\eqref{eq:cond_l1} is satisfied and the phases of \emph{all} nonzero entries of \inputvec are i.i.d. and uniformly distributed on $[0,2\pi)$, we can apply~\cite[Thm. 14]{tropp2008} (with $\delta=\inputdim^{-s}$) to infer that \inputvec is the unique solution of both (\Pzero) and (\Pone) with probability at least $(1-\inputdim^{-s})(1-2\inputdim^{-s})\geq(1-3\inputdim^{-s})$.
\end{IEEEproof}
\paragraph*{Interpretation of~\fref{thm:l0_l1_unique}}
We next present an interpretation of our result and reveal the fundamental property that allows to break the square-root bottleneck with reduced randomness.
In particular, we determine conditions on the dictionary such that both $\elemA=\scales(\outputdim/(\log\inputdim))$ and $\elemB=\scales(\outputdim/(\log\inputdim))$. As a consequence, a robust sparsity threshold $\sparsity>\elemA+\elemB=\scales(\outputdim/(\log\inputdim))$ is established.
In the following, for clarity of exposition, we only consider the dependency of \elemA and \elemB on the 
dictionary parameters \coh, \coha, \cohb, \inputdimA, \inputdimB, and the spectral norms of \dicta and \dictb, and absorb all constants that are independent of these quantities in $c(\gamma,s)$, where $\gamma$ and $s$ are defined in~\fref{thm:l0_l1_unique}. Note that $c(\gamma,s)$ can change its value at each appearance.
Condition~\eqref{eq:condA} together with $\elemA\leq\inputdimA$ yields the following constraint on \elemA:
\be
	\elemA\leq c(\gamma,s)\min\lefto\{\coh^{-2}/(\log\inputdim),\coha^{-1},{\inputdimA}\right\}.
\een
This constraint is compatible with $\elemA=\scales(\outputdim/(\log\inputdim))$, if the following three requirements are fulfilled:
\renewcommand{\labelenumi}{\roman{enumi})}
\begin{enumerate}
	\item the coherence of \dict satisfies $\coh=\order(1/\sqrt{\outputdim})$
	\item the coherence of  \dicta satisfies $\coha=\order((\log\inputdim)/\outputdim)$
	\item the cardinality of \dicta satisfies $\inputdimA\geq c\,\outputdim/(\log\inputdim)$
\end{enumerate}
where $c$ is a constant that can change at each appearance.
Condition~\eqref{eq:condB}, which can be rewritten as 
\be\label{eq:upperB}
	\elemB\leq c(\gamma,s)\min\lefto\{\frac{\cohb^{-2}}{\log\inputdim},\frac{\inputdimB}{\spectralnorm{\dictb}^2},\frac{\inputdimB}{\spectralnorm{\dicta}^2\spectralnorm{\dictb}^2}\right\}
\ee
is more laborious to interpret. For the constraint \eqref{eq:upperB} to be compatible with $\elemB=\scales(\outputdim/(\log\inputdim))$, we need requirement i) above to be fulfilled (recall that $\cohb\leq\coh$), together with the following two requirements on the spectral norms of \dictb and \dicta, namely
\begin{enumerate}
\setcounter{enumi}{3}
	\item $\spectralnorm{\dictb}^2\leq c\,\inputdimB(\log\inputdim)/\outputdim$
	\item $\spectralnorm{\dicta}^2\leq c\,\inputdimB(\log\inputdim)/(\spectralnorm{\dictb}^2\!\outputdim)$.
\end{enumerate}
We finally note that when the requirements i) -- v) are met, conditions~\eqref{eq:cond_l0} and~\eqref{eq:cond_l1}, which can then be rewritten as $\elemA+\elemB\leq c\,\outputdim$ and $\elemA+\elemB\leq c\,\outputdim/(\log\inputdim)$,
%
%
respectively, are compatible with both $\elemA=\scales(\outputdim/(\log\inputdim))$ and $\elemB=\scales(\outputdim/(\log\inputdim))$.

Hence, a robust sparsity threshold $\sparsity>\elemA+\elemB=\scales(\outputdim/(\log\inputdim))$ can be established under the same probabilistic model on \inputvec as in the two-ONB case; namely,
the positions of the nonzero entries of \inputvec corresponding to \dictb have to be chosen randomly, while the positions of the nonzero entries of \inputvec  corresponding to \dicta can be chosen arbitrarily. 
%
%

The requirements iv) and v) are difficult to interpret because they depend on the spectral norms of the sub-dictionaries \dicta and \dictb. To get more insight into these two requirements, we consider the special case of \dicta and \dictb being tight frames for $\complexset^\outputdim$~\cite{christensen03} (with the frame elements \elltwo-normalized to one). Then, $\spectralnorm{\dicta}^2=\inputdimA/\outputdim$ and $\spectralnorm{\dictb}^2=\inputdimB/\outputdim$, so that iv) is trivially satisfied and v) reduces to $\inputdimA\leq c\,\outputdim\log\inputdim$. However, because of the Welch lower bound~\cite{welch1974lower} condition ii) puts a more stringent restriction on the cardinality of \inputdimA for large \outputdim. Hence, a robust sparsity threshold of $\scales(\outputdim/(\log\inputdim))$ is obtained, under the same probabilistic model on the vector \inputvec as in the two-ONB case, if the coherence of sub-dictionary \dicta satisfies $\coha=\order((\log\inputdim)/\outputdim)$.
%

%

%

\paragraph*{A simple dictionary that satisfies i) - v)}
For $\outputdim=p^k$, with $p$ prime and $k\in\naturals^+$, a dictionary \dict with coherence equal to $1/\sqrt{\outputdim}$ can be obtained by concatenating $\outputdim+1$ ONBs for $\complexset^\outputdim$~\cite{gribonval2003}. Since \dict constitutes a tight frame for $\complexset^\outputdim$, by~\cite{tropp2008} a robust sparsity threshold of $\scales(\outputdim/(\log\inputdim))$ is obtained by randomizing over all positions of the nonzero entries of \inputvec.
Note, however, that we can write $\dict=[\dicta\,\,\dictb]$, where \dicta is an ONB ($\coha=0$) and \dictb  is the concatenation of the remaining $\outputdim$ ONBs and hence a tight frame for $\complexset^\outputdim$. As $\inputdimA=\outputdim$ the requirements iii) and v) are satisfied. Therefore, by the results of the previous paragraph, a robust sparsity threshold of $\scales(\outputdim/(\log\inputdim))$  is obtained by randomizing only over the positions of the nonzero entries of \inputvec corresponding to \dictb.


%
%
\appendices
\vspace{-0.1cm}
\section{Proof of \fref{lem:subdictionary}}\label{app:proofs}
%
%
%
Since the minimum singular value $\sigma_\text{min}(\subD)$ of the sub-dictionary \subD can be lower-bounded as $\sigma_\text{min}^2(\subD)\geq1-\spectralnorm{\subD^H\subD-\matI_{\elemA+\elemB}}$, we have 
\begin{multline}
\Prob\lefto\{\sigma_\text{min}(\subD)\leq 1/\sqrt{2}\right\}  = \Prob\lefto\{\sigma_\text{min}^2(\subD)\leq 1/2\right\} \\
	 \leq \Prob\lefto\{1-\spectralnorm{\subD^H\subD-\matI_{\elemA+\elemB}}\leq1/2\right\}\\
	 = \Prob\lefto\{\spectralnorm{\subD^H\subD-\matI_{\elemA+\elemB}}\geq 1/2\right\}.\label{eq:singularVSnorm}
\end{multline}
Next, we quantify the tail behavior of the random variable $\hollowS=\spectralnorm{\subD^H\subD-\matI_{\elemA+\elemB}}$, which will then lead to an upper bound on the probability of $\sigma_\text{min}(\subD)$ falling below $1/\sqrt{2}$. To this end the following lemma will be useful.
\begin{lem}[\!{\cite[Prop. 10]{tropp2008}}]\label{lem:tailbound}
If~the~mo\-ments of the nonnegative random variable $R$ can be upper-bounded as $[\Exop(R^q)]^{1/q}\leq\alpha\sqrt{q}+\beta$ for all $q\geq Q\in\integers_0^+$, where $\alpha,\beta\in\reals_0^+$, then, 
\be
	\Prob\{R\geq e^{1/4}(\alpha u+\beta)\}\leq e^{-u^2/4}
\een
for all $u\geq\sqrt{Q}$.
\end{lem}
\vspace{3mm}
To be able to apply~\fref{lem:tailbound} to $\hollowS=\spectralnorm{\subD^H\subD-\matI_{\elemA+\elemB}}$, we first need  an upper bound on $[\Exop(\hollowS^q)]^{1/q}$ that is of the form $\alpha\sqrt{q}+\beta$. We start by writing the sub-dictionary \subD  as $\subD = [\subA\,\,\subB]$,
%
where \subA and \subB denote the matrices containing the columns chosen arbitrarily from \dicta and randomly from \dictb, respectively.
We then obtain
\be
	\subD^H\subD-\matI_{\elemA+\elemB} = \lefto[\!\!\begin{array}{cc}\subA^H\subA-\matI_\elemA\!\!\!\! & \subA^H\subB\\ \subB^H\subA\!\!\!\! & \subB^H\subB-\matI_\elemB\end{array}\!\!\right].
\een
Applying the triangle inequality for operator norms, we can now upper-bound  $\hollowS$ according to
\ba
	\hollowS & = \spectralnorm{\!\lefto[\!\!\begin{array}{cc}\subA^H\subA-\matI_\elemA\!\!\!\! & \subA^H\subB\\ \subB^H\subA\!\!\!\! & \subB^H\subB-\matI_\elemB\end{array}\!\!\right]\!}\nonumber\\
	& \leq \spectralnorm{\!\lefto[\!\!\begin{array}{cc}\subA^H\subA-\matI_\elemA\!\!\!\! & \mathbf{0}\\ \mathbf{0} \!\!\!\!& \subB^H\subB-\matI_\elemB\end{array}\!\!\right]\!}+ \spectralnorm{\!\lefto[\!\!\begin{array}{cc}\mathbf{0} \!\!\!\!& \subA^H\subB\\ \subB^H\subA \!\!\!\!& \mathbf{0}\end{array}\!\!\right]\!}\nonumber\\
	%
	&\leq\max\lefto\{\spectralnorm{\subA^H\subA-\matI_\elemA},\spectralnorm{\subB^H\subB-\matI_\elemB}\right\}+\spectralnorm{\subA^H\subB}\nonumber\\
	&\leq\spectralnorm{\subA^H\subA-\matI_\elemA}+\spectralnorm{\subB^H\subB-\matI_\elemB}+\spectralnorm{\subA^H\subB}\label{eq:spectralsum}
\ea
%
%
where the second inequality follows because the spectral norm of both a block-diagonal matrix and an anti-block-diagonal matrix is given by the largest among the spectral norms of the individual nonzero blocks.
Next, we define $\hollowA=\spectralnorm{\subA^H\subA-\matI_\elemA}$, $\hollowB=\spectralnorm{\subB^H\subB-\matI_\elemB}$, and $\cross=\spectralnorm{\subA^H\subB}$.
It then follows from~\eqref{eq:spectralsum} that for all $q\geq 1$
\ba
	\lefto[\Exop (\hollowS^q)\right]^{1/q}&\leq\lefto[\Exop\lefto(\lefto(\hollowA+\hollowB+\cross\right)^q\right)\right]^{1/q}\nonumber\\
	&\leq \lefto[\Exop(\hollowA^q)\right]^{1/q}+\lefto[\Exop(\hollowB^q)\right]^{1/q}+\lefto[\Exop\lefto(\cross^q\right)\right]^{1/q}\nonumber\\
	&\leq\hollowA+\lefto[\Exop(\hollowB^q)\right]^{1/q}+\lefto[\Exop\lefto(\cross^q\right)\right]^{1/q}\label{eq:threeterms}
\ea
%
where the second inequality is a consequence of the triangle inequality for the norm $[\Exop(\abs{\cdot}^q)]^{1/q}$  (recall that $q\geq1$), and in the last step we used the fact that \hollowA is a deterministic quantity. All expectations in~\eqref{eq:threeterms} are with respect to the random choice of columns from sub-dictionary \dictb. 

We next upper-bound the three terms on the right-hand side (RHS) of \eqref{eq:threeterms} individually. Applying Ger\v{s}gorin's disc theorem~\cite[Th.~6.1.1]{hornjohnson} to the first term, we obtain
\be
	 \hollowA=\spectralnorm{\subA^H\subA-\matI_\elemA} \leq (\elemA-1)\coha.\label{eq:firstterm}
\ee
For the second term on the RHS of \eqref{eq:threeterms} we can use~\cite[Eq. 6.1]{tropp2008} to get 
\ba\label{eq:tropp6.1}
	\lefto[\Exop(\hollowB^q)\right]^{1/q}&=\lefto[\Exop\lefto(\spectralnorm{\subB^H\subB-\matI_\elemB}^q\right)\right]^{1/q}\nonumber\\
	&\leq\sqrt{144\cohb^2\elemB r_1}+2\elemB\spectralnorm{\dictb}^2\!/\inputdimB
\ea
where $r_1=\max\lefto\{1,\log\lefto(\elemB/2+1\right)\!,q/4\right\}$.
Assuming that $q\geq\max\{4\log(\elemB/2+1),4\}$ and hence $r_1=q/4$, we can simplify~\eqref{eq:tropp6.1} to
\be\label{eq:secondterm}
	\lefto[\Exop(\hollowB^q)\right]^{1/q}\leq6\sqrt{\cohb^2\elemB}\sqrt{q}+\frac{2\elemB}{\inputdimB}\spectralnorm{\dictb}^2.
\ee
To bound the third term on the RHS of~\eqref{eq:threeterms}, we use the upper bound on the spectral norm of a random compression~\cite[Thm. 8]{tropp2008} combined with $\rank(\subA^H\subB)\leq\elemB$. This yields
\ba
	\lefto[\Exop\lefto(\cross^q\right)\right]^{1/q}&=\lefto[\Exop\lefto(\spectralnorm{\subA^H\subB}^q\right)\right]^{1/q}\nonumber\\
	&\leq3\sqrt{r_2}\,\vecnorm{\subA^H\dictb}_{1,2}+\sqrt{\frac{\elemB}{\inputdimB}}\spectralnorm{\subA^H\dictb}\label{eq:lastterm1}
\ea
where $r_2=\max\lefto\{2,2\log\elemB,q/2\right\}$.
Assuming that $q\geq\max\{4\log\elemB,4\}$, we can further bound the RHS of~\eqref{eq:lastterm1} to get
\ba
	\lefto[\Exop\lefto(\cross^q\right)\right]^{1/q}&\leq\frac{3}{\sqrt{2}}\sqrt{q}\,\vecnorm{\subA^H\dictb}_{1,2}+\sqrt{\frac{\elemB}{\inputdimB}}\spectralnorm{\subA^H\dictb}\nonumber\\
	&\leq \frac{3}{\sqrt{2}}\sqrt{\coh^2\elemA}\sqrt{q}+\sqrt{\frac{\elemB}{\inputdimB}}\spectralnorm{\subA^H\dictb}\label{eq:lasttermcoh}\\
	&\leq\frac{3}{\sqrt{2}}\sqrt{\coh^2\elemA}\sqrt{q}+\sqrt{\frac{\elemB}{\inputdimB}}\spectralnorm{\dicta}\!\spectralnorm{\dictb}\label{eq:lasttermspectralnorm}
\ea
where~\eqref{eq:lasttermcoh} follows from the fact that the magnitude of each entry of $\subA^H\dictb$ is upper-bounded by $\coh$ and, thus, $\vecnorm{\subA^H\dictb}_{1,2}\leq \sqrt{\coh^2\elemA}$. To arrive at~\eqref{eq:lasttermspectralnorm} we used $\spectralnorm{\subA^H\dictb}\leq\spectralnorm{\subA^H}\!\spectralnorm{\dictb}\leq\spectralnorm{\dicta}\!\spectralnorm{\dictb}$, which follows from the sub-multiplicativity of the spectral norm and the fact that the spectral norm of the submatrix \subA of \dicta cannot exceed that of \dicta. 
We can now combine the upper bounds~\eqref{eq:firstterm},~\eqref{eq:secondterm}, and~\eqref{eq:lasttermspectralnorm} to obtain  
\ba
%
%
\lefto[\Exop\lefto(\hollowS^q\right)\right]^{1/q}&\leq(\elemA-1)\coha+6\sqrt{\cohb^2\elemB}\sqrt{q}+\frac{2\elemB}{\inputdimB}\spectralnorm{\dictb}^2+\nonumber\\
	&\dummyrel{=}\,\,+\frac{3}{\sqrt{2}}\sqrt{\coh^2\elemA}\sqrt{q}+\sqrt{\frac{\elemB}{\inputdimB}}\spectralnorm{\dicta}\!\spectralnorm{\dictb}\nonumber\\
	%
	& = \underbrace{\lefto(6\sqrt{\cohb^2\elemB}+3\sqrt{\coh^2\elemA/2}\right)}_{\alpha}\sqrt{q}\,+\nonumber\\[-0.1cm]
	&\dummyrel{=}\,\,+\underbrace{(\elemA-1)\coha+\frac{2\elemB}{\inputdimB}\spectralnorm{\dictb}^2+\sqrt{\frac{\elemB}{\inputdimB}}\spectralnorm{\dicta}\!\spectralnorm{\dictb}}_{\beta}\nonumber\\[-0.2cm]
	& = \alpha\sqrt{q}+\beta\nonumber
%
%
\ea
for all $q\geq Q_1=\max\{4\log(\elemB/2+1),4\log\elemB,4\}$.
Hence,~\fref{lem:tailbound} yields
\be
	\Prob\{\hollowS\geq e^{1/4}(\alpha u+\beta)\}\leq e^{-u^2/4}
\een
for all $u\geq\!\sqrt{Q_1}$. In particular, under the assumption $\inputdim\geq e\approx2.7$, it follows that the choice $u=\sqrt{4s\log\inputdim}$ satisfies $u\geq\!\!\sqrt{Q_1}$ for any $s\geq1$. Straightforward calculations reveal that conditions~\eqref{eq:condA} and~\eqref{eq:condB} ensure that $e^{1/4}(\alpha u+\beta)\leq1/2$, which together with~\eqref{eq:singularVSnorm} then leads to
\ba
	\Prob\lefto\{\sigma_\text{min}(\subD)\leq 1/\sqrt{2}\right\}&\leq \Prob\lefto\{\hollowS\geq1/2\right\} \nonumber\\
	&\leq \Prob\{\hollowS\geq e^{1/4}(\alpha u+\beta)\}\nonumber\\
	&\leq e^{-u^2/4} = \inputdim^{-s}.\nonumber\\[-0.6cm]
	\nonumber
\ea
%
%
%
\renewcommand{\baselinestretch}{0.98}\small\normalsize
\bibliography{IEEEabrv,confs-jrnls,publishers,patrick}
\bibliographystyle{IEEEtran}

\end{document}